\documentclass[11pt,fleqn]{article}
\usepackage{amssymb,latexsym,amsmath,amsfonts,graphicx, ebezier}

    \advance\textheight by \topskip

    \numberwithin{equation}{section}

    \def\dbar{{\bar\partial \,}}

    \def\bigO{{\cal O}}

    \def\sech{{\rm sech}}

    \def\P2n{{\rm P}_{{\rm II}}^{(n)}}

    \newtheorem{theorem}{Theorem}[section]

    \newtheorem{Definition}[theorem]{Definition}
    
    \newtheorem{Remark}[theorem]{Remark}
    \newenvironment{remark}{\begin{Remark}\rm}{\end{Remark}}
    \newtheorem{Example}[theorem]{Example}
    
    \newtheorem{Assumptions}[theorem]{Assumptions}

    \newcommand{\e}{\epsilon}
\newcommand{\lb}{\lambda}

    {\rm \trivlist \item[\hskip \labelsep{\bf Proof. }]}%
    {\hspace*{\fill}$\Box$\endtrivlist}
    {\rm \trivlist \item[\hskip \labelsep{\bf Proof}]}%
    {\hspace*{\fill}$\Box$\endtrivlist}

\begin{document}
\title{The Riemann-Hilbert approach to obtain critical asymptotics for Hamiltonian perturbations of hyperbolic and elliptic systems}
\author{Tom Claeys}

\maketitle

\begin{abstract}
The present paper gives an overview of the recent developments in the description of critical behavior for Hamiltonian perturbations of hyperbolic and elliptic systems of partial differential equations. It was conjectured that this behavior can be described in terms of distinguished Painlev\'e transcendents, which are universal in the sense that they are, to some extent, independent of the equation and the initial data.
We will consider several examples of well-known integrable equations that are expected to show this type of Painlev\'e behavior near critical points. The Riemann-Hilbert method is a useful tool to obtain rigorous results for such equations. We will explain the main lines of this method
 and we will discuss the universality conjecture from a Riemann-Hilbert point of view.
\end{abstract}

\section{Introduction}

We will discuss the asymptotic behavior of solutions to
Hamiltonian perturbations of hyperbolic and elliptic systems of
partial differential equations near critical points of the
unperturbed system. We will give an overview of recent developments
in this area using the Riemann-Hilbert (RH) approach. The systematic
study of the critical behavior of such Hamiltonian systems was
initiated by Dubrovin and collaborators \cite{Dubrovin, Dubrovin2,
Dubrovin3, Dubrovin4, DGK} in a series of papers where it was
conjectured that the local behavior of solutions near critical
points is to a large extent universal, and that it can be described
in terms of distinguished Painlev\'e transcendents. The conjectured behavior has been proved and verified numerically in various
special cases.
We will focus mainly on particular cases of Hamiltonian PDEs or
systems of PDEs that can be written in a so-called Lax form and that
consequently can be solved using the direct and inverse scattering transform,
which can be formulated as a RH problem. The Deift/Zhou steepest
descent method then provides the tools to obtain asymptotics for the
RH problem and for the solution to the system of
equations.

\medskip

\begin{figure}[t]
\begin{center}
\includegraphics[scale=0.5]{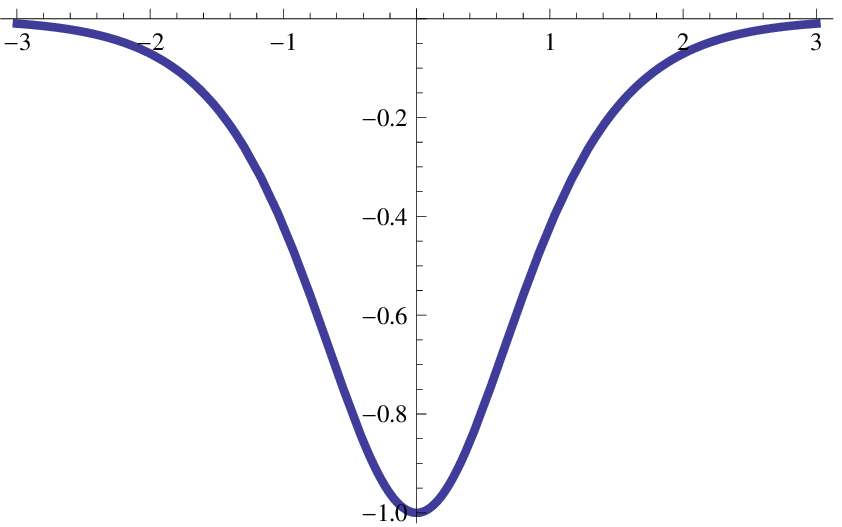}
\includegraphics[scale=0.5]{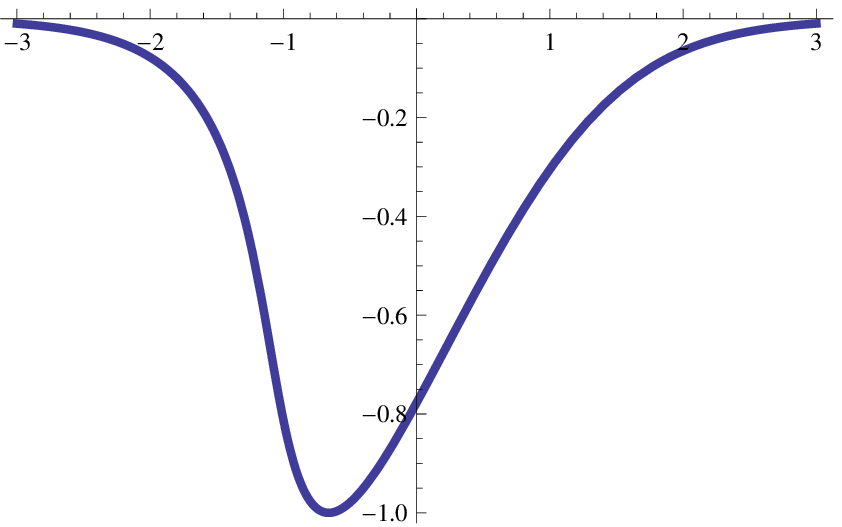}
\includegraphics[scale=0.5]{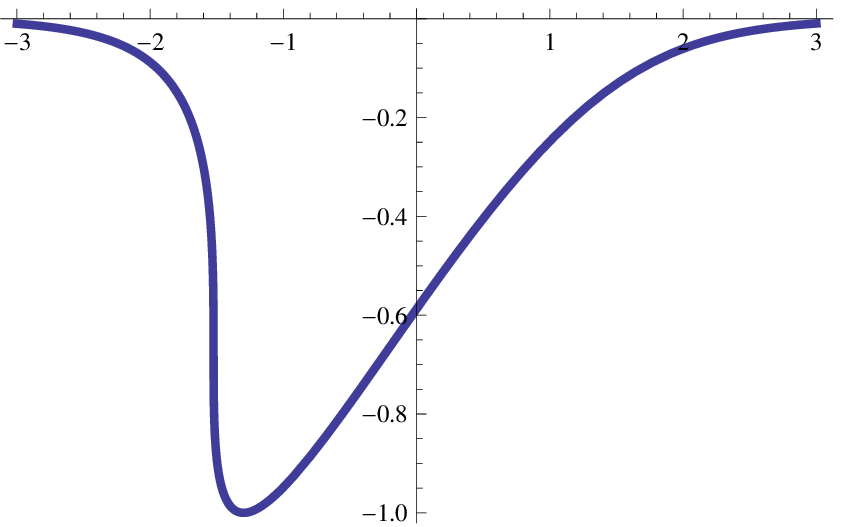}
\end{center}
\caption{The solution to the Hopf equation with initial data $u_0(x)=-\sech^2(x)$ for $t=0$, $t=0.11$, and $t=0.216506$.} \label{figure: Hopf}
\end{figure}

Let us first consider the simple quasi-linear hyperbolic equation
$u_t+6uu_x=0$, which is known as the Hopf equation or in-viscid
Burgers' equation. It is a classical fact that solutions to this
equation exist only for small times $t>0$ and develop shocks at
a certain time. Given initial data $u_0(x)$, the solution is given
by the method of characteristics in the implicit form
\begin{equation}\label{Hopf} u(x,t)=u_0(\xi),\qquad x=6tu_0(\xi)+\xi.
\end{equation}
For simplicity, we will only consider smooth initial data that decay
sufficiently rapidly at $\pm\infty$. One observes that the
derivative of the solution blows up at the critical time
\[
t_c=\dfrac{1}{\max_{\xi\in\mathbb{R}}[-6u'_0(\xi)]},
\]
see Figure \ref{figure: Hopf}. We write $x_c$ for the point where
the derivative of $u(.,t_c)$ blows up, and $u_c=u(x_c,t_c)$.

The blow-up of the first derivative is called the gradient
catastrophe, and we say that $x_c$ and $t_c$ are the point and time
of gradient catastrophe. Locally near $x_c$, the solution at the
critical time behaves (for generic initial data) like
\begin{equation}\label{Hopf local}
u(x,t_c)= u_c-c(x-x_c)^{1/3}+\bigO(x-x_c)^{4/3},\qquad \mbox{as
$x\to x_c$.}
\end{equation}

In order to avoid the gradient catastrophe, one can perturb the Hopf
equation. The Burgers' equation $u_t+6uu_x-\e^2u_{xx}=0$ is an
example of a dissipative regularization, but we will concentrate on
dispersive or Hamiltonian perturbations of the Hopf equation. In
\cite{Dubrovin}, all Hamiltonian perturbations to the Hopf equation
$u_t+6uu_x=0$ up to order $4$ have been classified, which lead to
the class of equations
\begin{multline}
\label{Hampert} u_t +6uu_x + \frac{\epsilon^2}{24} \left[ 2 c
u_{xxx} + 4 c' u_x u_{xx}
+ c'' u_x^3\right]+\epsilon^4 \left[ 2 p u_{xxxxx} \right.\\
\left. +2 p'( 5 u_{xx} u_{xxx} + 3 u_x u_{xxxx}) + p''( 7 u_x
u_{xx}^2 + 6 u_x^2 u_{xxx} ) +2 p''' u_x^3 u_{xx}\right]=0,
\end{multline}
where $p=p(u)$ and $c=c(u)$ are arbitrary smooth functions. For
$c=12$, $p=0$, we have the KdV equation
  \begin{equation}\label{KdV}u_t+6uu_x+\e^2u_{xxx}=0.\end{equation}
  For sufficiently smooth initial data $u_0(x)=u(x,0,\e)$,
  solutions to the KdV equation exist for all times $t>0$. For small
  times, $u$ and its derivatives are bounded and the third
  derivative
  term will only contribute up to order $\bigO(\e^2)$ for small
  $\e$.
 Near the time and point of gradient catastrophe for the Hopf equation, solutions to KdV start forming
 oscillations, and the $x$-derivative of the solution is
 no longer uniformly bounded. The contribution of the dispersive term
 $\e^2u_{xxx}$ in the equation becomes visible in this region: for $t$ slightly bigger than $t_c$,
 it is well-known that there is
an interval $[x^-(t),x^+(t)]$ where the KdV solution
 $u(x,t,\e)$ shows oscillatory behavior. Outside this interval, the
 KdV solution can still be approximated by a continuation of the
 Hopf solution.
The oscillations have been studied extensively in many works and are
given asymptotically in the small dispersion limit in terms of the
solutions to the Whitham equations for KdV and in terms of Jacobi
elliptic functions and elliptic integrals \cite{DVZ, DVZ2, GP, LL, FRT,
V2}.

At least before the gradient catastrophe, the small $\e$ behavior
for general equations in the family (\ref{Hampert}) is expected to
be similar to the behavior for KdV, but few analytical results are
available. Even existence of solutions with smooth initial data for
small times has not been established in general. For small times, it
is not hard to believe that the contributions of the terms of order
$\e^2$ and $\e^4$ will be small. One expects that
$u(x,t,\e)=u(x,t)+\bigO(\e^2)$ for $t<t_c-\delta$, where $u(x,t)$ is
the solution to the Hopf equation with initial data $u_0(x)$. It is
less obvious what happens for $x$ and $t$ near the point and time of
gradient catastrophe. This problem was addressed by Dubrovin in
\cite{Dubrovin}, where universality of critical behavior for
Hamiltonian perturbations of hyperbolic PDEs was conjectured.

\subsection*{Universality conjecture: the hyperbolic case}
It is expected that a generic solution to an equation of the
form (\ref{Hampert}) has an expansion of the form
\begin{equation}\label{expansionu}
u(x,t,\e)=u_c+c_1\e^{2/7}U(c_2\frac{x-x_c-c_3(t-t_c)}{\e^{\frac{6}{7}}},c_4\frac{t-t_c}{\e^{4/7}})+\bigO(\e^{4/7}),
\end{equation}
as $\e\to 0$ and at the same time $x\to x_c$ and $t\to t_c$ at
appropriate speeds, where $x_c$ and $t_c$ are the point and time of gradient catastrophe. The constants $c_1, c_2,
c_3, c_4$ depend on the equation and the initial data, but not on
$x,t$. The function $U(X,T)$ is expected to be universal, i.e.\
independent of the equation and independent of the choice of initial
data. It should be a solution to the fourth order ODE
\begin{equation}\label{PI2}
        X=TU-\left(\frac{1}{6}U^3+\frac{1}{24}(U_X^2+2UU_{XX})
            +\frac{1}{240}U_{XXXX}\right)
    \end{equation} which is smooth and real for all real values of $X$ and
    $T$. The existence and uniqueness of such a solution was part of
    the conjecture, but the existence of a real smooth solution with the asymptotic behavior
\begin{equation}\label{as U}
U(X,T)=\mp (6|X|)^{1/3}+\bigO(|X|^{-2/3}),\qquad \mbox{ as
$X\to\infty$},
\end{equation}
for any fixed $T\in\mathbb R$, has been proved in \cite{CV1}.
The smoothness of the Painlev\'e transcendent $U(X,T)$ was already conjectured long time ago for $T=0$ \cite{BMP, Moore}, and appeared later to be related to the Gurevich-Pitaevskii special solution to KdV \cite{GP, Suleimanov1, Suleimanov2}. In the physics literature, $U(X,T)$ shows up in the study of ideal incompressible liquids \cite{KS1, KS2, GST} and in quantum gravity \cite{DSS}.
It appears also in the description of the local
eigenvalue behavior in unitary random matrix ensembles with singular
edge points \cite{BMP, CV2}.

 Substituting the asymptotics (\ref{as U}) in
(\ref{expansionu}) for $t=t_c$, we recover (\ref{Hopf local}) in the
limit $\e\to 0$. Equation (\ref{PI2}) is known as the second member
of the Painlev\'e I hierarchy and has, given $T\in\mathbb C$,
solutions that are meromorphic in $X$ with an infinite number of
complex poles. The smoothness of $U$ means in other words that no
poles lie on the real line. It is remarkable that $U(X,T)$ is an exact solution to the KdV equation written in the form $U_T+UU_X+\frac{1}{12}U_{XXX}=0$.

Loosely speaking, the conjecture suggests that the local behavior of
solutions to Hamiltonian perturbations of the Hopf equation is
universally described in terms of $U(X,T)$: the same Painlev\'e
transcendent appears independent of the initial data and independent
of the equation. $U$ can be seen as the function which describes the
transition between the region where the small dispersion asymptotics
are determined by the Hopf equation and the region of oscillatory
behavior.

Critical behavior in terms of $U(X,T)$ is expected \cite{Dubrovin2} for
 more general Hamiltonian
perturbations of strictly hyperbolic systems of equations of the
form
\begin{equation}\label{Hampertsyst}
u_t+A(u)u_x=0,\qquad u=\begin{pmatrix}u_1\\u_2\\ \vdots \\
u_n\end{pmatrix},
\end{equation}
where $A(u)$ is a $n\times n$ matrix-valued function of $u$, strict hyperbolicity means that $A$ has
real and distinct eigenvalues in the domain of the $(x,t)$-plane
under consideration.

In the case of the KdV equation and KdV hierarchy (see below), the
universality conjecture has been proved for a class of negative, real analytic
initial data $u_0(x)$ with one local minimum, and which tend to $0$
rapidly as $x\to\pm\infty$. For the KdV equation, we have
(\ref{expansionu}) with the values \cite{CG1}
\[
c_1=\frac{2}{(8k)^{2/7}} ,\quad c_2=\frac{1}{(8k)^{1/7}},\quad
c_3=6u_c,\quad c_4=\frac{12}{(8k)^{3/7}},\] where $k=-f_L'''(u_c)$,
with $f_L$ the inverse of the decreasing part of the initial data
$u_0$. The expansion holds in a neighborhood of $(x_c,t_c)$ which
shrinks with $\e$, to be more precise it holds in the double scaling
limit where $\e\to 0$, $x\to x_c$, $t\to t_c$ in such a way that
\begin{equation}
\lim c_2\frac{x-x_c-c_3(t-t_c)}{\e^{\frac{6}{7}}}=X,\qquad \lim
c_4\frac{t-t_c}{\e^{4/7}}=T,\qquad X,T\in\mathbb R,
\end{equation}

\subsection*{Examples}
We will now list a number of equations and systems of equations that are expected to belong to the universality class for which $U(X,T)$ describes the solutions locally near the gradient catastrophe.
\begin{itemize}
\item[(i)] The $m$-th time flow of the KdV hierarchy is given by
\begin{equation}
u_t-(-1)^{m}\partial_x\mathcal L_m=0,
\end{equation}
where $\mathcal L_m$ is the Lenard-Magri recursion operator defined
by
\begin{equation}
\partial_x\mathcal
L_m=\left(\e^2\partial_x^3+4u\partial_x+2u_x\right)\mathcal
L_{m-1},\qquad \mathcal L_0=u.
\end{equation}
For $m=1$, we have the KdV equation, and
if we impose $\mathcal L_m$ to be zero for $u=0$, the second and third equation in the hierarchy are given by
\begin{align}
\label{KdV2} &u_{t} - 30 u^2 u_x - \epsilon^2 \left(20u_x u_{xx} +
10
u u_{xxx}\right) - \epsilon^4 u_{xxxxx} = 0,\\
 &u_{t}
+140u^3u_x+\e^2\left(70u_x^3+280uu_xu_{xx}+70u^2u_{3x}\right)\nonumber\\
&\qquad\qquad\qquad\qquad\qquad\quad+\e^4\left(70
u_{2x}u_{3x}+42u_xu_{4x}+ 14uu_{5x}\right)+\e^6u_{7x}=0\label{KdV3},
\end{align}
where we have written $u_{jx}$ for the $j$-th partial derivative of $u$ with respect to $x$.
The universality conjecture has been confirmed numerically for the
second member of the hierarchy in \cite{DGK2}, and an expansion of
the form (\ref{expansionu}) was proved analytically in \cite{CG4}
for all equations in the hierarchy.
\item[(ii)] For the Camassa-Holm (CH) equation
\begin{equation}\label{CH}u_t+(3u+2\kappa)u_x-\e^2\left(u_{xxt}+2u_xu_{xx}+uu_{xxx}\right)=0,\qquad t>0,\ x\in\mathbb
R,\ \kappa\in\mathbb R,
\end{equation}
numerical results have been obtained in \cite{GK} supporting the
universality conjecture, although the equation is not precisely of the
form (\ref{Hampert}) or (\ref{Hampertsyst}), see \cite{GK1}.
\item[(iii)] The de-focusing nonlinear Schr\"odinger (NLS) equation
\begin{equation}i\e\psi_t+\frac{\e^2}{2}\psi_{xx}-|\psi|^2\psi=0,\label{NLS df}\end{equation} can be
transformed to the system
\begin{equation}\label{system NLS}
\begin{cases}u_t+vu_x+uv_x=0\\
v_t+vv_x-u_x+\frac{\e^2}{4}\left(\frac{1}{2}\frac{u_x^2}{u^2}-\frac{u_{xx}}{u}\right)_x=0\end{cases},
\end{equation}
where $u=-|\psi|^2$,
$v=\frac{\e}{2i}\left(\frac{\psi_x}{\psi}-\frac{\psi_x^*}{\psi^*}\right)$.
This is a Hamiltonian perturbation of the system
\begin{equation}\label{unpert NLS}
\begin{pmatrix}
u\\v
\end{pmatrix}_t+\begin{pmatrix}v&u\\-1&v\end{pmatrix}\begin{pmatrix}
u\\v
\end{pmatrix}_x=0,
\end{equation}
which is hyperbolic because the eigenvalues of the coefficient
matrix are $v\pm\sqrt{-u}$.
\item[(iv)] For the Kawahara equation
\begin{equation}
u_t+6uu_x+\e^2u_{xxx}=\e^4u_{xxxxx}
\end{equation}
and generalized KdV equations
\begin{equation}
u_t+6u^nu_x+\e^2u_{xxx}=0,
\end{equation}
 the universality conjecture has been
verified numerically in \cite{DGK2}.
\end{itemize}

\subsection*{Universality conjecture: the elliptic case} The
focusing nonlinear Schr\"odinger equation
\begin{equation}i\e\psi_t+\frac{\e^2}{2}\psi_{xx}+|\psi|^2\psi=0\label{NLS f}\end{equation} can be seen as a Hamiltonian
perturbation of an elliptic system, since it is transformed to the
system (\ref{system NLS}) after the substitutions $u=|\psi|^2$,
$v=\frac{\e}{2i}\left(\frac{\psi_x}{\psi}-\frac{\psi_x^*}{\psi^*}\right)$.
This is a Hamiltonian perturbation of the system (\ref{unpert NLS})
which is then of elliptic type since the eigenvalues $v\pm
i\sqrt{u}$ are complex conjugate for positive $u$. The Cauchy
problem for this equation is ill-posed and the blow-up phenomenon is
essentially different than for hyperbolic systems. The system
(\ref{unpert NLS}) can be solved using a hodograph transform method
up to a critical time $t_c$ where the spatial derivative of $u$
blows up at a point $x_c$. This happens when the peak at a local
maximum of the solution becomes more and more narrow (or focusses).
This critical point is called a point of elliptic umbilic
catastrophe, and a conjecture has been formulated in \cite{DGK}
concerning the local behavior of the NLS equation near this point.
The key role in this conjecture is played by the Boutroux
tritronqu\'ee solution to the first Painlev\'e equation
\begin{equation}
Q_{ZZ}=Z+6Q^2,
\end{equation}
characterized by the asymptotic behavior
\begin{equation}\label{as Q}
Q(Z)\sim-\sqrt{\frac{Z}{6}},\qquad\mbox{ as $Z\to\pm\infty$ in the
sector $|\arg Z|<\frac{4\pi}{5}$.}
\end{equation}
It is known that this is a meromorphic function with infinitely many
poles in the sector of the complex plane where $|\arg Z|\geq
\frac{4\pi}{5}$. In the complementary sector $|\arg Z|<
\frac{4\pi}{5}$, $Q$ has no poles for sufficiently large $Z$ (this
would contradict the asymptotic behavior (\ref{as Q})) and no poles
at all on the positive half-line \cite{JK}, but this does not
exclude the possibility that a finite number of poles is present in
this sector. It was conjectured by Dubrovin, Grava, and Klein in
\cite{DGK} that $Q$ has no poles at all in the sector $|\arg Z|<
\frac{4\pi}{5}$. The second part of the conjecture, in analogy to
the hyperbolic case, describes the behavior of the solution to the
focusing NLS equation: it is expected to have an asymptotic
expansion where the first error term to the unperturbed solution is
of order $\e^{2/5}$ and can be described completely in terms of the
tritronqu\'ee solution $Q(Z)$ evaluated at complex arguments $Z$.

Substantial progress has been made on a proof of the second part of
the conjecture in the recent work \cite{BT}, we will comment in more
detail on this later on. The poles of $Q$ have been studied in
\cite{Masoero}, but the first part of the conjecture is still open.
It was discussed in \cite{Dubrovin2} that an expansion in terms of
$Q(Z)$ is not exclusively expected for the focusing NLS equation,
but also for more general Hamiltonian perturbations of elliptic
systems. The tritronqu\'ee solution $Q$ is also relevant for the
asymptotics of recurrence coefficients of certain orthogonal
polynomials on a complex contour \cite{DK, FokasItsKitaev}.

\subsection*{Outline} We will not discuss classification results or
try to formulate the most general form of the conjectures, for these
topics we refer to \cite{Dubrovin, Dubrovin2, Dubrovin3, Dubrovin4,
DGK, DGK2}. Rather we want to focus on techniques that can be used
to prove the universality conjectures in various special cases. The
best one can do so far is to try to prove the universality
conjectures for systems of equations that can be solved using the
inverse scattering transform. Then one can formulate a RH problem
which characterizes solutions to the system of equations, and an
asymptotic analysis of this RH problem can in principle lead to the
critical asymptotics in terms of the Painlev\'e transcendents $U$
and $Q$, which themselves can be characterized in terms of a
so-called model RH problem. Without going into the technical details, we will give an overview of this method
in the case of the KdV equation (following \cite{DVZ, DVZ2, CG1}),
and indicate the differences for the KdV hierarchy (following
\cite{CG4}). Afterwards we will recall the RH problems associated to
the Camassa-Holm equation \cite{BdMS} and the de-focusing and
focusing NLS equation \cite{Shabat}, and explain the main lines of
the procedure that can be followed to obtain asymptotics for
solutions of those equations. We will also give heuristic arguments
supporting the universal nature of the critical behavior from the
Riemann-Hilbert point of view.

\section{RH problems for Painlev\'e transcendents}
The Painlev\'e transcendents $U$ and $Q$, which are the central
objects in the universality conjectures, can be characterized in
terms of RH problems. We will state the RH problems here, and we
will explain that the absence of poles is equivalent to the
solvability of the RH problems.

\subsection{RH problem for $P_I^2$ solution $U$}\label{section: Painleve} The
function $U(X,T)$ is characterized by the following RH problem.
\subsubsection*{RH problem for $\Psi$:}
\begin{itemize}
    \item[(a)] $\Psi=\Psi(\zeta;X,T)$ is analytic for $\zeta\in\mathbb{C}\setminus\Gamma$,
    with $\Gamma=\mathbb R\cup e^{\frac{6\pi i}{7}}\mathbb R^+\cup e^{-\frac{6\pi i}{7}}\mathbb R^+$ oriented
    as in Figure
    \ref{figure: contour gamma}.
    \item[(b)] $\Psi$ has continuous boundary values $\Psi_\pm$ ($+$ referring to the left side of the contour, $-$ to the right side according to
    the orientation in Figure \ref{figure: contour gamma}) on
    $\Gamma\setminus\{0\}$, and we have
    \begin{align}
        \label{RHP Psi: b1}
        &\Psi_+(\zeta)=\Psi_-(\zeta)
        \begin{pmatrix}
            0 & 1 \\
            -1 & 0
        \end{pmatrix},& \mbox{for $\zeta\in\mathbb R^-$,} \\[1ex]
        \label{RHP Psi: b2}
        &\Psi_+(\zeta)=\Psi_-(\zeta)
        \begin{pmatrix}
            1 & 1 \\
            0 & 1
        \end{pmatrix},& \mbox{for $\zeta\in \mathbb R^+$,} \\[1ex]
        \label{RHP Psi: b3}
        &\Psi_+(\zeta)=\Psi_-(\zeta)
        \begin{pmatrix}
            1 & 0 \\
            1 & 1
        \end{pmatrix},& \mbox{for $\zeta\in e^{\pm \frac{6\pi i}{7}}\mathbb R^+$.}
    \end{align}
    \item[(c)] $\Psi$ has the following behavior at infinity,
    \begin{equation}\label{RHP Psi: c}
        \Psi(\zeta)=\left(I+\frac{A_1(X,T)}{\zeta}+\bigO(\zeta^{-2})\right)\zeta^{-\frac{1}{4}\sigma_3}N
        e^{-\theta(\zeta;X,T)\sigma_3},
    \end{equation}
    where $\sigma_3=\begin{pmatrix}1&0\\0&-1\end{pmatrix}$, and $N$ and $\theta$ are given by
    \begin{equation}\label{def N theta}
    N=\frac{1}{\sqrt 2}\begin{pmatrix}1&1\\-1&1\end{pmatrix}e^{-\frac{\pi
    i}{4}\sigma_3}, \qquad
    \theta(\zeta;X,T)=\frac{1}{105}\zeta^{7/2}-\frac{T}{3}\zeta^{3/2}+X\zeta^{1/2},
    \end{equation}
and where $A_1(X,T)$ is a matrix independent of $\zeta$.
\end{itemize}
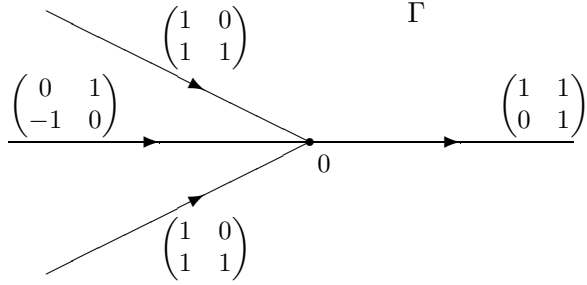
\begin{figure}[t]
    \begin{center}
    \setlength{\unitlength}{1mm}
    \begin{picture}(95,47)(0,2)
        \put(30,38){\small $\begin{pmatrix}1&0\\1&1\end{pmatrix}$}
        \put(10,29){\small $\begin{pmatrix}0&1\\-1&0\end{pmatrix}$}
        \put(30,10){\small $\begin{pmatrix}1&0\\1&1\end{pmatrix}$}
        \put(75,29){\small $\begin{pmatrix}1&1\\0&1\end{pmatrix}$}

        \put(63,41){$\Gamma$}

        \put(50,25){\thicklines\circle*{.9}}
        \put(51,21){\small 0}

        \put(50,25){\line(-2,1){35}} \put(36,32){\thicklines\vector(2,-1){.0001}}
        \put(50,25){\line(-2,-1){35}} \put(36,18){\thicklines\vector(2,1){.0001}}
        \put(50,25){\line(-1,0){40}} \put(30,25){\thicklines\vector(1,0){.0001}}
        \put(50,25){\line(1,0){35}} \put(70,25){\thicklines\vector(1,0){.0001}}
    \end{picture}
    \caption{The jump contour $\Gamma$ and the jump matrices for $\Psi$}
        \label{figure: contour gamma}
    \end{center}
\end{figure}
It was proved in \cite{CV1} that this RH problem has a solution for
all real values of $X,T$, and that the function
\begin{equation}
U(X,T)=2A_{1,11}(X,T)-A_{1,12}^2(X,T)
\end{equation}
is a real pole-free solution to equation (\ref{PI2}) which has the
asymptotic behavior (\ref{as U}). General solutions to (\ref{PI2})
are characterized by a similar RH problem, but with jumps also on
the four rays $e^{\pm \frac{2\pi i}{7}}\mathbb R^+$ and $e^{\pm
\frac{4\pi i}{7}}\mathbb R^+$ \cite{Kapaev}. The jump matrices are
all triangular with ones on the diagonal, and the triangular
structure is as follows: the jump matrices are upper-triangular on
$\mathbb R$, $e^{\pm \frac{4\pi i}{7}}\mathbb R^+$, and
lower-triangular on $e^{\pm \frac{2\pi i}{7}}\mathbb R^+$ and
$e^{\pm \frac{6\pi i}{7}}\mathbb R^+$. The off-diagonal entries of
the jump matrices are called Stokes multipliers and satisfy certain
constraints. Given $T\in\mathbb C$, there is a one-to-one
correspondence between the set of admissible Stokes multipliers
(also called the monodromy surface) and the solutions of the $P_I^2$
equation. The RH problem is known to be solvable if and only if the
corresponding Painlev\'e solution is analytic at the value $(X,T)$,
or in other words the RH problem is not solvable if and only if
$(X,T)$ is a pole of the corresponding Painlev\'e solution.

\subsection{RH problem for PI}
The Boutroux tritronqu\'ee solution $Q(Z)$ of the Painlev\'e I
equation is characterized by another RH problem.

\subsubsection*{RH problem for $\Phi$:}
\begin{figure}[t]
    \begin{center}
    \setlength{\unitlength}{1mm}
    \begin{picture}(95,47)(0,2)
        \put(45,38){\small $\begin{pmatrix}1&i\\0&1\end{pmatrix}$}
        \put(10,29){\small $\begin{pmatrix}0&i\\i&0\end{pmatrix}$}
        \put(45,11){\small $\begin{pmatrix}1&i\\0&1\end{pmatrix}$}
        \put(75,29){\small $\begin{pmatrix}1&0\\i&1\end{pmatrix}$}


        \put(50,25){\thicklines\circle*{.9}}
        \put(48,26){\small 0}

        \put(50,25){\line(2,3){15}} \put(60,40){\thicklines\vector(2,3){.0001}}
        \put(50,25){\line(2,-3){15}} \put(60,10){\thicklines\vector(2,-3){.0001}}
        \put(50,25){\line(-1,0){40}} \put(30,25){\thicklines\vector(1,0){.0001}}
        \put(50,25){\line(1,0){35}} \put(70,25){\thicklines\vector(1,0){.0001}}
    \end{picture}
    \caption{The jump contour $\Gamma'$ and the jump matrices for $\Phi$}
    \end{center}
    \label{figure: contour phi}
\end{figure}
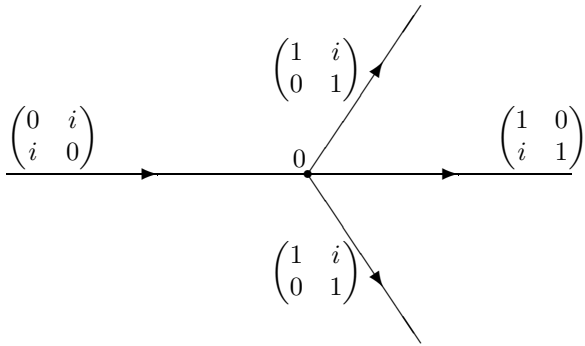

\begin{itemize}
    \item[(a)] $\Phi=\Phi(\zeta;Z)$ is analytic for $\zeta\in\mathbb{C}\setminus\Gamma'$,
    with $\Gamma'=\mathbb R\cup e^{\frac{2\pi i}{5}}\mathbb R^+\cup e^{-\frac{2\pi i}{5}}\mathbb R^+$ oriented
    as in Figure
    \ref{figure: contour phi}.
    \item[(b)] $\Phi$ satisfies the following jump relations on
    $\Gamma'\setminus\{0\}$,
    \begin{align}
        \label{RHP Phi: b1}
        &\Phi_+(\zeta)=\Phi_-(\zeta)
        \begin{pmatrix}
            0 & i \\
            i & 0
        \end{pmatrix},& \mbox{for $\zeta\in\mathbb R^-$,} \\[1ex]
        \label{RHP Phi: b2}
        &\Phi_+(\zeta)=\Phi_-(\zeta)
        \begin{pmatrix}
            1 & 0 \\
            i & 1
        \end{pmatrix},& \mbox{for $\zeta\in \mathbb R^+$,} \\[1ex]
        \label{RHP Phi: b3}
        &\Phi_+(\zeta)=\Phi_-(\zeta)
        \begin{pmatrix}
            1 & i \\
            0 & 1
        \end{pmatrix},& \mbox{for $\zeta\in e^{\pm i\frac{2\pi}{5}}\mathbb R^+$.}    \end{align}
    \item[(c)] $\Phi$ has the following behavior at infinity,
    \begin{equation}\label{RHP Phi: c}
        \Phi(\zeta)=\left(I+\frac{B_1(Z)}{\zeta}+\bigO(\zeta^{-2})\right)\zeta^{-\frac{1}{4}\sigma_3}
        \frac{1}{\sqrt 2}\begin{pmatrix}1&1\\1&-1\end{pmatrix}
        e^{-\tilde\alpha(\zeta;Z)\sigma_3},
    \end{equation}
    where $\tilde\alpha$ is given by
    \begin{equation}\label{def rho}
    \tilde\alpha(\zeta;Z)=\frac{4}{5}\zeta^{5/2}-Z\zeta^{1/2}.
    \end{equation}
\end{itemize}
The function
\begin{equation}
Q(Z)=B_{1,21}^2-2B_{1,22}
\end{equation}
is the tritronqu\'ee solution to Painlev\'e I \cite{Kapaev2}, and
given $Z\in\mathbb C$, the RH problem for $\Phi$ is solvable if and
only if $Z$ is not a pole of $Q$. A general solution to the
Painlev\'e I equation can be obtained in terms of a RH problem with
triangular jumps on the contour $\mathbb R\cup e^{\pm
i\frac{2\pi}{5}}\mathbb R^+\cup e^{\pm i\frac{4\pi}{5}}\mathbb R^+$
\cite{Kapaev2, FIKN}.

In order to discuss the solvability of this RH problem, let us first
introduce the {\em homogeneous RH problem for $\Phi$}: we say that
$\Phi_0$ is a solution to the homogeneous RH problem if it satisfies
conditions (a) and (b) of the RH problem for $\Phi$, and it has the
asymptotic behavior
\begin{equation}
\Phi_0(\zeta)=\bigO(\zeta^{-3/4})e^{-\tilde\alpha(\zeta;Z)\sigma_3},\qquad\mbox{
as $\zeta\to \infty$.}\end{equation} Following the vanishing lemma
procedure \cite{FIKN, FokasMuganZhou, FokasZhou}, the solvability of
the RH problem for $\Phi$ is equivalent to the fact that the
homogeneous version of the RH problem has only the trivial zero
solution: the RH problem for $\Phi$ with parameter $Z\in\mathbb C$
is solvable if and only if the homogeneous RH problem has only the
vanishing solution $\Phi_0\equiv 0$. To prove this, one can follow
the procedure developed in \cite{DKMVZ2, FokasMuganZhou, FokasZhou}
(which works for a fairly general class of RH problems) and show
that a certain singular integral operator related to the RH problem
is a Fredholm operator of index zero.

So in order to have smoothness of $Q$ for $|\arg Z|<\frac{4\pi}{5}$,
it suffices to show that the homogeneous RH problem has only the
zero solution (such a result is called a vanishing lemma) for $|\arg Z|<\frac{4\pi}{5}$. Then the RH problem for
$\Phi$ is solvable, and $Q$ has no pole at $Z$.

 A vanishing lemma has been proved for the $P_I^2$ RH problem for real $X$ and $T$ \cite{CV1}, but this turns out to be
considerably more complicated in the PI case, even for positive real
values of $Z$.

\section{Small dispersion asymptotics for solutions to KdV: the Riemann-Hilbert approach}
We will now give an overview of the RH method to obtain
asymptotics for solutions to the KdV equation in the small
dispersion limit. The method consists of three steps: first one has
to construct a RH problem that characterizes solutions to the KdV
equation. This can be done using scattering theory for the
Schr\"odinger operator. The jump matrices in the RH problem will
depend on the initial data via a reflection coefficient. The second
step requires asymptotic information about the reflection
coefficient as $\e\to 0$. Such results were obtained in
\cite{Ramond} using WKB techniques. Once this information is
gathered, one can start with the final step which is the asymptotic
analysis of the RH problem using the Deift/Zhou steepest descent
method \cite{DZ1}.

\subsection{Construction of a RH problem related to the Schr\"odinger
operator}\label{section: RHP KdV} The key observation here is that
the KdV equation can be written in the Lax form $L_t=[L,A]$, where
$L$ is the Schr\"odinger operator, and $A$ is an antisymmetric
operator:
\begin{equation}\label{LA}Lf =\e^2\partial_x^2f + u f, \qquad A f =
A=4\epsilon^2\partial_x^3 f+3\left(u\partial_x
f+\partial_x(uf)\right).\end{equation}

We consider potentials $u(x)$ that are smooth and decay sufficiently
fast at $\pm\infty$. For a general real potential $u(x)$, $L$ has a
finite number of purely imaginary eigenvalues $\lambda_j$ for which
the linear Schr\"odinger equation $Lf=\lambda f$ has solutions
decaying both at $+\infty$ and at $-\infty$. For $\lambda<0$, it
admits solutions $\psi_\pm$ and $\phi_\pm$ satisfying the
oscillatory asymptotic conditions
\begin{align}
\label{normalization1}
&\lim_{x\to +\infty}\psi_\pm(\lb;x,t,\epsilon) e^{\pm\frac{i}{\e}\sqrt{-\lb} x}=1,\\
&\label{normalization2}\lim_{x\to -\infty}\phi_\pm(\lb;x,t,\epsilon)
^{\mp\frac{i}{\e}\sqrt{-\lb}x}=1.
\end{align}
Those solutions are called Jost solutions. They are related by a
connection matrix given by
\begin{equation}
\label{transition}
\begin{pmatrix}\psi_+(\lb)&\psi_-(\lb)\end{pmatrix}=\begin{pmatrix}\phi_-(\lb)&\phi_+(\lb)\end{pmatrix}
\begin{pmatrix} a(\lb;t,\e)& \overline b(\lb;t,\e)\\
 b(\lb;t,\e)& \overline a(\lb;t,\e)
\end{pmatrix},\qquad \lb<0.
\end{equation}
The ratio $r(\lambda;t,\e)=\frac{b}{a}(\lambda;t,\e)$ is called the
reflection coefficient for the Schr\"odinger equation.

It is well-known that $\psi_+$ ($\psi_-$) and $\phi_+$ ($\phi_-$)
can be extended analytically to the upper (lower) half plane. The
vector-valued function
\begin{equation}\label{M}
M(\lb;x,t,\e)=\begin{cases}\begin{pmatrix}\phi_+&\frac{1}{a}
\psi_+\end{pmatrix}e^{\frac{-i}{\e}x(-\lambda)^{1/2}\sigma_3},&
\mbox{ as $\lambda\in\mathbb C^+$},\\[3ex]
\begin{pmatrix}\frac{1}{a^*}\psi_-&\phi_-\end{pmatrix}e^{\frac{-i}{\e}x(-\lambda)^{1/2}\sigma_3},&\mbox{
as $\lb\in\mathbb C^-$},
\end{cases}
\end{equation}
is meromorphic in $\mathbb C\setminus R$ and it can be showed using
(\ref{transition}) and asymptotic properties of the Jost solutions
that $M$ satisfies the following RH conditions \cite{Shabat, DT,
DVZ, CG1}
 \subsubsection*{RH problem for $M$}
\begin{itemize}
\item[(a)] $M(\lb;x,t,\e)$ is meromorphic for
$\lb\in\mathbb{C}\backslash \mathbb{R}$, \item[(b)] $M$ has
continuous boundary values $M_\pm(\lambda)$ for $\lambda\in\mathbb
R\setminus\{0\}$ that satisfy the jump conditions
\begin{align}&\label{RHP M1}M_+(\lb)=M_-(\lb){\small \begin{pmatrix}1&
r(\lb;t,\e)e^{\frac{2i}{\e}x(-\lambda)^{1/2}}\\-\bar{r}(\lb;t,\e)
e^{\frac{-2i}{\e}x(-\lambda)^{1/2}} &1-|r(\lb;t,\e)|^2
\end{pmatrix}}&\mbox{ for $\lb<0$,}\\
&M_+(\lb)=M_-(\lb)\sigma_1,\quad
\sigma_1=\begin{pmatrix}0&1\\1&0\end{pmatrix}&\mbox{ for $\lb>0$}.
\end{align}
\item[(c)] As $\lambda\to\infty$, we have $M(\lambda;x,t,\e)\to
\begin{pmatrix}1&1\end{pmatrix}$.
\end{itemize}
The potential $u(x,t,\e)$ can then be recovered from the equation
\begin{equation}\label{uM}
u(x,t,\e)=-2i\e\lim_{\lambda\to\infty}
\sqrt{-\lambda}\partial_x(M_{11}(\lambda;x,t,\e)-1).
\end{equation}
In general $M$ has a finite number of simple poles on the imaginary
axis, which are the eigenvalues of the Schr\"odinger operator. One
has to impose additional conditions on the residues of $M$ to have a
unique RH solution. If the potential $u(x)$ is negative, the
Schr\"odinger equation has no point spectrum and $M$ is holomorphic
in $\mathbb C\setminus \mathbb R$. Then one can also use the
$x$-derivatives of the Jost solutions to construct the second row of
a $2\times 2$ matrix $M$ satisfying the same jump condition, and the
asymptotic condition
\begin{equation}
M(\lambda)\sim\begin{pmatrix}1&1\\i\sqrt{-\lambda}&-i\sqrt{-\lambda}\end{pmatrix},\qquad\mbox{ as $\lambda\to\infty$.}
\end{equation}
A matrix RH problem is for several reasons more convenient to work
with than a vector RH problem.

The above construction works for any potential $u(x)$ decaying
sufficiently fast at $\pm\infty$, it is not necessary that
$u=u(x,t,\e)$ solves the KdV equation. However, for general
potentials $u(x,t,\e)$, the time-dependence of $r$ is not explicitly
given. On the other hand if $u(x,t,\e)$ satisfies a suitable Lax
type equation $L_t=[L,A]$, the time evolution of $r$ can be
explicitly calculated, and the eigenvalues $\lambda_j$ are
independent of $t$. In the KdV case, one has \cite{GGKM}
\begin{equation}
r(\lambda;t,\e)=r_0(\lambda;\e)e^{\frac{8i}{\e}t(-\lb)^{\frac{3}{2}}}.
\end{equation}
This is a crucial observation: if we know the initial reflection
coefficient corresponding to the initial data $u_0(x)$ (the direct
scattering transform), then we know the reflection coefficient at
any time $t>0$. In order to find the solution $u(x,t,\e)$ at time
$t>0$, it suffices to solve the RH problem for $M$ at time $t>0$
(the inverse scattering transform).


\subsection{Semi-classical asymptotics for the reflection coefficient}
The solution to the Cauchy problem for KdV is characterized in terms
of the RH problem for $M$. The jump matrix for $M$ depends on the
reflection coefficient $r_0$, which depends in a simple explicit way
on $t$, but in a much more complicated way on the spectral variable
$\lambda$ and on $\e$. Since we are interested in the small $\e$
behavior of the RH solution $M$, inevitably we need small $\e$
asymptotics for the initial reflection coefficient $r_0$. Detailed
asymptotics were obtained in \cite{Ramond} for negative initial data
with a single negative hump that are analytic in a sufficiently
large neighborhood of the real line, and we will concentrate on such
data in what follows. For $\lambda<-1$, $r_0(\lambda;\e)$ decays
rapidly as $\e\to 0$, and for $-1<\lambda<0$, it is oscillatory and
has the leading order behavior
\begin{equation}\label{as r0}
r_0(\lambda;\e)\sim i\exp\left(-\frac{2i}{\e}\rho(\lambda)\right),
\qquad\mbox{ as $\e\to 0$},
\end{equation}
where $\rho(\lambda)=\frac{1}{2}\int_\lambda^0
\frac{f_L(\xi)}{\sqrt{\xi-\lambda}}d\xi$, and $f_L$ is the inverse
of the decreasing part of the initial data $u_0$. Near $-1$, there
is a subtle transition between the decay and oscillatory behavior.

Substituting $r_0$ in the jump matrix of the RH problem by its
leading order asymptotics (\ref{as r0}), we obtain the jump relation
$M_+(\lambda)=M_-(\lambda)v_M(\lambda)$ for $\lambda\in\mathbb R$,
with
\begin{align}&\label{RHP M1-2}v_M(\lb)\sim I &\mbox{ for $\lb<-1$,}\\
&v_M(\lb)\sim {\small
\begin{pmatrix}1&i\exp\left(-\frac{2i}{\e}(\rho(\lambda)-\alpha(\lambda))\right)
\\i\exp\left(\frac{2i}{\e}(\rho(\lambda)-\alpha(\lambda))\right) &0
\end{pmatrix} }&\mbox{ for $-1<\lb<0$,}\\
&v_M(\lb)= \sigma_1&\mbox{ for $\lb>0$},
\end{align}
where $\alpha(\lambda)=x(-\lambda)^{1/2}+4t(-\lambda)^{3/2}$. It is not obvious at all that one can proceed
with the RH analysis, neglecting the corrections to the leading
order behavior of the reflection coefficient. This needs to be
justified, and especially for $\lambda$ near $-1$, technical
issues need to be handled. We will not address this here, since it
is our goal to give a flavour of the method only, we refer the
reader to \cite{CG1, CG4} for details.

\subsection{Steepest descent analysis of the RH problem}

The Deift/Zhou steepest descent method was developed in \cite{DZ1}
and applied to the small dispersion limit for the KdV equation in \cite{DVZ, DVZ2}. The method
consists of a number of invertible transformations of the RH problem,
and the goal of this series of transformations is to obtain a RH
problem for which the jump matrices are uniformly close to the
identity matrix as $\e\to 0$ (or in a more complicated double
scaling limit where $x\to x_c$, $t\to t_c$ simultaneously with $\e\to 0$), and with a solution that tends to $I$ as $\lambda\to
\infty$. For such RH problems, one can apply small norm theory to
prove that the solution is uniformly close to the identity matrix for small $\e$
and to obtain the next terms in the asymptotic series of the
solution. Inverting each of the transformations, one obtains
asymptotics for $M$ from the asymptotics for the final small norm RH
solution.

The first transformation requires the construction of a
$g$-function, which has suitable jump conditions and asymptotic
behavior. For $x,t$ sufficiently close to $x_c,t_c$, the $g$-function $g=g(\lambda;x,t)$ is defined by the scalar RH conditions
\begin{itemize}
\item[(a)] $g$ is analytic in $\mathbb C\setminus [u,+\infty)$,
\item[(b)] $g$ has the jump relations
\begin{align}
&\label{prop g1}g_{+}(\lambda)+g_{-}(\lambda)=0,&\mbox{ for $\lambda\in(0,+\infty)$,}\\
&\label{prop g} g_{+}(\lambda)+
g_{-}(\lb)-2\rho(\lambda)+2\alpha(\lambda)=0,&\mbox{ for
$\lambda\in(u,0)$},
\end{align}
\item[(c)] $g(\lambda)=\bigO(\lambda^{-1/2})$ as $\lambda\to\infty$.
\end{itemize}
Given $u$, this scalar RH problem has a solution (which is unique if
one imposes suitable behavior of $g$ near $u$ and $0$) that can be
expressed explicitly as an integral transform of the inverse of the
initial data. In order to analyze the KdV solution in the vicinity
of $x_c$ and $t_c$, the choice of the point $u$ is crucial: it has to be chosen as
$u=u_c=u(x_c,t_c)$.

\begin{figure}[t]
\begin{center}
    \setlength{\unitlength}{1.4mm}
    \begin{picture}(137.5,26)(22,11.5)
        \put(112,25){\thicklines\circle*{.8}}
        \put(45,25){\thicklines\circle*{.8}}
        \put(47,26){\small $-1-\delta$}
        \put(112,26){\small $0$}\put(122,27){\small $\sigma_1$}
        \put(85,25){\thicklines\circle*{.8}} \put(80,26){\small $u_c$}
        \put(99,25){\thicklines\vector(1,0){.0001}}
        \put(85,25){\line(1,0){45}}
        \put(124,25){\thicklines\vector(1,0){.0001}}
        \put(22,25){\line(1,0){23}}
        \put(35,25){\thicklines\vector(1,0){.0001}}
\put(74,13){\small $\begin{pmatrix}1&0\\i
e^{-\frac{2i}{\epsilon}\phi}&1\end{pmatrix}$}
        \qbezier(45,25)(65,45)(85,25) \put(66,35){\thicklines\vector(1,0){.0001}}
        \qbezier(45,25)(65,5)(85,25) \put(66,15){\thicklines\vector(1,0){.0001}}
\put(73,36){\small $\begin{pmatrix}1&i
e^{\frac{2i}{\epsilon}\phi}\\0&1\end{pmatrix}$} \put(34,27){\small
$I$} \put(90,29){\small $\begin{pmatrix}
e^{-\frac{2i}{\epsilon}\phi_+}&i\\i& 0
\end{pmatrix}$}
 \put(49,32){$\Sigma_1$}
\put(49,16){$\Sigma_2$}
    \end{picture}
    \caption{The jump contour $\Sigma_S$ and the jumps for $S$ in the limit $\e\to 0$}
    \label{figure: sigmaS}
\end{center}
\end{figure}
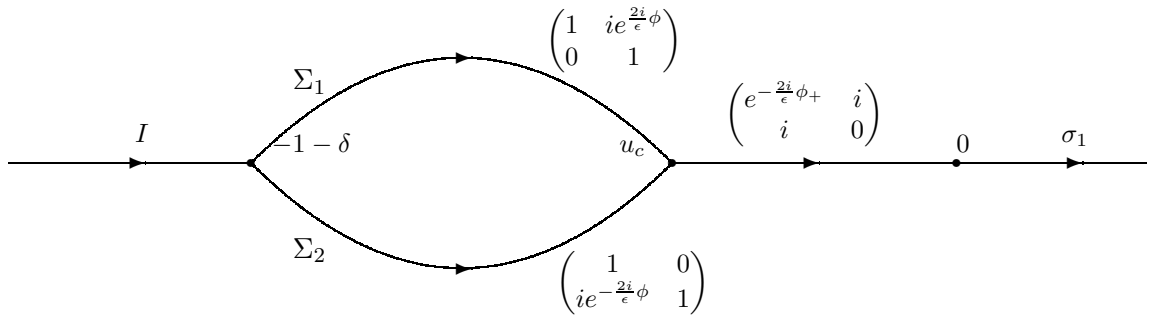

The $g$-function is an essential ingredient for the asymptotic
analysis of the RH problem for $M$. It clears the road for a
transformation of the RH problem to one with jump conditions of the
following form, on a jump contour as the one shown in Figure
\ref{figure: sigmaS}:

\subsubsection*{RH problem for $S$}
\begin{itemize}
\item[(a)] $S$ is analytic in $\mathbb C\setminus \Sigma_S$,
\item[(b)] $S_+(\lambda)=S_-(\lambda)v_S$ for $\lambda\in\Sigma_S$, and as $\e\to 0$, the jump matrix $v_S$ has the leading order behavior
\begin{equation}\label{vS}
v_S(\lambda)\sim \begin{cases}
\begin{array}{lr}
\begin{pmatrix}1& ie^{\frac{2i}{\e}\phi(\lb)}\\
0&1
\end{pmatrix},&\mbox{ on $\Sigma_1$},\\[3ex]
 \begin{pmatrix}1&0\\
ie^{-\frac{2i}{\e}\phi(\lb)}&1
\end{pmatrix},&\mbox{ on $\Sigma_2$,}\\[3ex]
\begin{pmatrix}e^{-\frac{2i}{\e}\phi_+(\lb)}
&i\\
i&0
\end{pmatrix},&\mbox{ as $\lambda\in (u_c,0)$,}\\[3ex]
I,&\mbox{ as
$\lambda\in(-\infty,-1-\delta_1)\cup(0,+\infty)$.}
\end{array}
\end{cases}
\end{equation}
\item[(c)] $S(\lambda)=\left(I+\bigO(\lambda^{-1})\right) \begin{pmatrix}1&1\\
i\sqrt{-\lambda}&-i\sqrt{-\lambda}\end{pmatrix}$ as
$\lambda\to\infty$.
\end{itemize}
The function $\phi$ appearing in the jump matrices is given by
\begin{equation}
\phi(\lambda)=g(\lambda)-\rho(\lambda)+\alpha(\lambda),
\end{equation}
and vanishes at the point $\lambda=u_c$. It can be expressed
explicitly as
\begin{multline}
\phi(\lambda;x,t)=-\sqrt{u_c-\lambda}
F(u_c;x,t)+\dfrac{2}{3}(u_c-\lambda)^{\frac{3}{2}}F'(u_c;x,t)\\-\dfrac{4}{15}(u_c-\lambda)^{\frac{5}{2}}F''(u_c;x,t)
-\dfrac{4}{15}\int_{u_c}^{\lambda}F'''(\xi;x,t)(\xi-\lambda)^{\frac{5}{2}}d\xi,\label{phi1}
\end{multline}
where $F(\lambda;x,t)=-x+6\lambda t+f_L(\lambda)$.

The sign of the real part of $\phi$ is such that the
exponentials in the jump matrices decay uniformly as $\e\to 0$ on
the jump contour, except in the vicinity of $u_c$, where $\phi$
behaves like
\begin{equation}\label{phi}
\phi(\lambda;x_c,t_c)=-c(u_c-\lambda)^{7/2}(1+\bigO(\lambda-u_c)),\qquad\mbox{
as $\lambda\to u_c$.}
\end{equation}
The constant $c$ is strictly positive whenever the generic
condition $f_L'''(u_c)\neq 0$ holds.

\subsubsection*{Global parametrix}
Ignoring a small neighborhood of $u_c$ and the corrections to the leading order behavior as $\e\to 0$, one arrives at a RH problem with jump only on $(u_c,+\infty)$: the jump on $(u_c,0)$ is $i\sigma_1$ and the jump on $(0,+\infty)$ is $\sigma_1$, where $\sigma_1=\begin{pmatrix}0&1\\1&0\end{pmatrix}$. This RH problem can be solved explicitly, indeed the function
\begin{equation}
\label{def Pinfty}
P^{(\infty)}(\lambda)=(-\lambda)^{1/4}(u_c-\lambda)^{-\sigma_3
/4}\begin{pmatrix}1&1\\ i& -i
\end{pmatrix}
\end{equation} satisfies the required jump relations and as $\lambda\to\infty$, it behaves as prescribed by condition (c) in the RH problem
for $S$.

\subsubsection*{Local parametrix}
Since we have neglected a neighborhood of $u_c$ so far, where the
jump matrix $v_S$ is not uniformly close to the identity matrix, it
is necessary to construct a local parametrix in the vicinity of
$u_c$, which satisfies approximately the same jump conditions than
$S$, and which 'matches' in an appropriate way with the global
parametrix. This construction is rather technical but the parametrix
can be constructed explicitly in terms of $\Psi$, the solution to
the $P_I^2$ RH problem discussed in Section \ref{section: Painleve}.
It appears that this is true independently of the precise form of
the function $\phi$. Only the local behavior of $\phi$ near $u_c$
matters: whenever $\phi(\lambda)\sim -c(u_c-\lambda)^{7/2}$ with
$c>0$ as $\lambda\to u_c$, the local parametrix can be constructed
in terms of $\Psi$. The power of vanishing $7/2$ can be identified
with the power $7/2$ in the asymptotic behavior for $\Psi$ in
(\ref{RHP Psi: c}) and (\ref{def N theta}). This is a first
indication that the $P_I^2$ solution $U(X,T)$ plays a role in much
more general settings. Indeed if we would consider the RH problem
for $M$, but with more complicated time-dependence of the reflection
coefficient (for example corresponding to another equation than
KdV), the local parametrix would still be built out of $\Psi$. This
supports, from the RH point of view, the conjecture that $U(X,T)$ is
in some sense universal.

\begin{remark}
If one wants to perform an asymptotic analysis of the RH problem for
$M$ away from the critical point $(x_c,t_c)$ as in \cite{DVZ} (for example for $t<t_c$), the
main difference is that $\phi(\lb)\sim -c(u-\lambda)^{3/2}$, so
$\phi$ vanishes at a slower rate. The local parametrix can in this
case be constructed in terms of the Airy function.
\end{remark}

\subsubsection*{Asymptotics for $M$}
Multiplying $S$ by the inverse of the local parametrix near $u_c$
and by the inverse of the global parametrix elsewhere, one obtains a
function $R$ which solves a so-called small-norm RH problem: one can
show that $R$ tends to the identity matrix in the double scaling
limit where $\e\to 0$, $x\to x_c$, $t\to t_c$ at appropriate speeds.
One can even obtain the next terms in the asymptotic series for $R$,
which are of order $\e^{1/7}$ and order $\e^{2/7}$, and which can be expressed explicitly in terms of $U(X,T)$. Inverting the
transformations $S\mapsto R$, $T\mapsto S$, and $M\mapsto T$, this
produces an asymptotic expansion for $M$ and, by (\ref{uM}), for the
KdV solution $u$. In particular the method described here allows one to prove that the KdV solution has an expansion of the form (\ref{expansionu}).

\begin{remark}
The above RH analysis works fine for initial data that are such that
the initial reflection coefficient is analytic and decays
sufficiently fast at infinity. If the initial reflection coefficient
would be $C^1$ or $C^2$ instead of analytic, the usual RH techniques
cannot be applied, but one may hope that recently developed dbar
techniques \cite{DMcL, McLM} could help to get around this problem.
\end{remark}

\section{The Riemann-Hilbert approach for other equations}
\subsection{The KdV hierarchy}
The equations in the KdV hierarchy can be written as a Lax equation
\begin{equation}\label{LaxKdVhierarchy}
L_{t}=[L,A_m],
\end{equation}
where $L$ is (as for the KdV equation) the Schr\"odinger operator
$L=\e^2\partial_x^2+u$ and where $A_m$ is an antisymmetric higher
order operator with leading order
$(-1)^{m+1}4^{m}\e^{2m}\partial_x^{2m+1}$. The lower order terms of
$A_m$ are determined by the requirement that $[L,A_m]$ is an
operator of multiplication with a function depending on $u, u_x,
\ldots, u_{(2m+1)x}$. For $m=1$, the operator $A$ was given in
(\ref{LA}). As explained in Section \ref{section:
RHP KdV}, the Jost solutions to the Schr\"odinger equation can be used to construct a RH problem. The only difference here is the time
evolution of the reflection coefficient, which is now given by
\begin{equation}
r(\lambda;t,\e)=r_0(\lambda;\e)e^{\frac{2.4^m
i}{\e}t(-\lb)^{\frac{2m+1}{2}}}.
\end{equation}
With this modification, the RH problem for $M$ given in Section
\ref{section: RHP KdV} still characterizes the solution to the
Cauchy problem for the $m$-th equation in the KdV hierarchy which is
given by (\ref{uM}). In the asymptotic analysis of the RH problem,
the $g$-function has to be modified and the function $\phi$ will be
somewhat different, but the main issue is that it will still behave
like in (\ref{phi}) locally near $u_c$. Details can be found in
\cite{CG4}.

\subsection{The Camassa-Holm equation}
The Camassa-Holm equation (\ref{CH}) is the compatibility condition
of the Lax equations
\begin{align}
&\label{Psix CH}\e^2f_{xx}=\left(\frac{1}{4}-\lambda(u-\e^2u_{xx}+\kappa)\right)f\\
&f_t=-\left(\frac{1}{2\lambda}+u\right)f_x+\frac{u_x}{2}f.
\end{align}
This fact has been used in \cite{BdMS} to construct a RH problem
that characterizes solutions to the Camassa-Holm equation. The RH
solution is, in a somewhat similar manner as for the KdV problem,
constructed using fundamental solutions to the spatial Lax equation
(\ref{Psix CH}), and the jump matrix contains a reflection
coefficient for this equation. Equation (\ref{Psix CH}) is related
to the Schr\"odinger equation through the unitary Liouville
transform \cite{BdMKST}. We present the RH problem in the variable
$z$, where $z^2=-\lambda$.

\subsubsection*{RH problem for $M^{CH}$}
\begin{itemize}
\item[(a)] $M^{CH}:\mathbb C\setminus \mathbb R\to \mathbb C^{1\times 2}$ is meromorphic for
$z\in\mathbb{C}\backslash \mathbb{R}$, \item[(b)] $M^{CH}$ has
continuous boundary conditions $M_\pm(\lambda)$ for
$z\in\mathbb R$ that satisfy the jump conditions
\begin{equation}\label{CH RHP M1}M_+^{CH}(z)=M_-^{CH}(z){\small \begin{pmatrix}1-|r(z;t,\e)|^2&
r(z;t,\e)e^{-2iyz}\\ -\bar{r}(z;t,\e)e^{2izy}
&1
\end{pmatrix}}\qquad\mbox{ for $z\in\mathbb R$,}
\end{equation}
\item[(c)] as $z\to\infty$, we have
\begin{equation}\label{CH RHP M:c}M^{CH}(z)=
\begin{pmatrix}1&1\end{pmatrix}+o(1).
\end{equation}
\end{itemize}
Again, in general there is a finite number of poles where one has to
add residue conditions to ensure the uniqueness of a RH solution.
In the simplest case there are no eigenvalues and $M^{CH}=M^{CH}(z;y,t,\e)$ is
holomorphic in $\mathbb C\setminus\mathbb R$.

The CH solution $u(x,t,\e)$ is recovered from the relation
\begin{equation}
u(x,t,\e)=\left.\frac{\partial}{\partial t}\ln
\frac{M_1^{CH}}{M_2^{CH}}(\frac{i}{2};y,t,\e)\right|_{y=y(x,t)},
\end{equation}
where $y(x,t)$ is given by
\[x=y+\ln\frac{M_1^{CH}}{M_2^{CH}}(\frac{i}{2};y,t,\e).\]
The time evolution for the reflection coefficient is explicitly
given:
\begin{equation}
r(z;t,\e)=r_0(z;\e)\exp\left(\frac{4i t z}{\e}\frac{1}{1+4z^2}\right).
\end{equation}
This RH problem has been used to obtain large time asymptotics for
CH solutions \cite{BdMKST, BdMS}. In the small dispersion setting,
there is an additional complication because the reflection
coefficient depends in a nontrivial way on the small parameter $\e$.
If one makes the ansatz that the initial reflection coefficient is analytic in $z$ and
behaves like
\begin{equation}\label{r0ansatz}
r_0(z;\e)\sim i\exp\left(-\frac{2i}{\e}\rho(z)\right),
\qquad\mbox{ as $\e\to 0$}
\end{equation}
for a suitable function $\rho$, one could proceed in a similar
manner as for the KdV RH problem, with the construction of a
$g$-function and the opening of the lens. For an appropriate choice of $u$, the $g$-function should
satisfy the conditions
\begin{itemize}
\item[(a)] $g$ is analytic in $\mathbb C\setminus [-u,u]$,
\item[(b)] $g$ has the jump relations
\[g_{+}(z)+
g_{-}(z)-2\rho(z)+2\beta(z;y,t)=0,\qquad\mbox{ for
$\lambda\in(-u,u)$},\]
\item[(c)] $g(z)\to 0$ as $z\to\infty$.
\end{itemize}
Here $\beta$ is given by
$\beta(z;y,t)=-z\left(y-\frac{2}{1+4z^2}t\right)$. After the
construction of the $g$-function, the lens can be opened as in the
KdV case. The jump matrices are then supposed to tend to the
identity matrix except near $\pm u$. One expects that the analogue
of $\phi$ will behave like $c|u\mp z|^{7/2}$ near $\pm u$ for
$x=x_c$, $t=t_c$, and that the local parametrix can be constructed
in terms of the RH problem for $\Psi$, but many technical issues
have to be solved in order to proceed with this method. Another
nontrivial problem one has to overcome to prove rigorously the
universality conjecture for Camassa-Holm, is to identify a class of
initial data such that $r_0$ is analytic and such that
(\ref{r0ansatz}) holds.

\subsection{The de-focusing NLS equation}
The NLS equation (\ref{system NLS}) is the compatibility
condition of the Lax pair equations
\begin{equation}\Psi_x=A\Psi,\qquad \Psi_t=B\Psi,\end{equation}
where
\begin{align}
&A=\frac{1}{\e}\begin{pmatrix}-i z&u\\ \overline{u}&i z
\end{pmatrix},\\
&B=\frac{1}{\e}\begin{pmatrix}-2i z^2-i|u|^2&2 z u+iu_x\\2 z
\overline{u}-i\overline{u}_x &2i z^2+i|u|^2.
\end{pmatrix}
\end{align}
 The spatial Lax equation is the Zakharov-Shabat equation,
and a scattering theory for it has been developed in \cite{BDT, Shabat, Zhou}.
Using special solutions to it, one constructs a solution to the
following matrix RH problem.

\subsubsection*{RH problem for $M^{NLS}$}
\begin{itemize}
\item[(a)] $M^{NLS}:\mathbb C\setminus \mathbb R\to \mathbb C^{2\times 2}$ is meromorphic for
$ z\in\mathbb{C}\backslash \mathbb{R}$, \item[(b)] $M^{NLS}$ has
continuous boundary conditions $M_\pm^{NLS}( z)$ for
$ z\in\mathbb R$ that satisfy the jump conditions
\begin{equation}\label{NLS RHP M1-2}M_+^{NLS}( z)=M_-^{NLS}( z){\small \begin{pmatrix}1-|r( z;t,\e)|^2&
-\bar r( z;t,\e)e^{-\frac{2i}{\e}x z}\\
{r}( z;t,\e)e^{\frac{2i}{\e}x z} &1
\end{pmatrix}}\qquad\mbox{ for $ z\in\mathbb R$,}
\end{equation}
\item[(c)] as $ z\to\infty$, we have
\begin{equation}\label{NLS RHP M:c}M^{NLS}( z;x,t,\e)=I+\bigO( z^{-1}).
\end{equation}
\end{itemize}
The NLS solution $u(x,t,\e)$ is recovered by
\begin{equation}
u(x,t,\e)=-2\lim_{ z\to \infty} z M_{12}^{NLS}( z;x,t,\e)
\end{equation}
The reflection coefficient $r( z;t,\e)$ has a time evolution
given by
\begin{equation}
r( z;t,\e)=r_0( z;\e)e^{\frac{4i}{\e}t z^2}.
\end{equation}
If one makes the ansatz $r_0( z;\e)\sim
e^{-\frac{2i}{\e}f_0( z)}$ for a suitable analytic function $f_0$, one
can start performing a Deift/Zhou steepest descent analysis for this
RH problem that should be somewhat similar to the analysis for the KdV
equation. An important question is then to find a class of $\e$-independent initial data for the de-focusing NLS equation that correspond to such reflection coefficients.

\subsection{The focusing NLS equation}
The RH problem for the focusing NLS equation is the same as for
de-focusing NLS, but with the modified jump relation
\begin{equation}\label{dfNLS RHP M1-2}M_+^{NLS}(z)=M_-^{NLS}(z){\small \begin{pmatrix}1+|r( z;t,\e)|^2&
\bar r( z;t,\e)e^{-\frac{2i}{\e}x z}\\
{r}( z;t,\e)e^{\frac{2i}{\e}x z} &1
\end{pmatrix}}\qquad\mbox{ for $ z\in\mathbb R$.}
\end{equation}
Although this may seem to be a minor change compared to the
de-focusing case, it has important consequences. Small dispersion asymptotics for the focusing NLS equation away from critical points have been studied for example in \cite{KMM, TV, TVZ}. In \cite{BT}, critical behavior was studied under the assumption that the initial reflection
coefficient is of the form $r_0( z)=e^{-\frac{2i}{\e} f_0( z)}$,
where $f_0$ needs to satisfy some assumptions. Because of the sign changes in
the jump matrix, one cannot construct a suitable $g$-function which
is analytic except on an interval of the real line, instead $g$ will
have its branch cut on a curve in the complex plane. Near two
special points $\alpha_1, \alpha_2$, the analogue of the function
$\phi$ we described in the KdV case behaves like $\phi(z)\sim c(
z-\alpha_j)^{5/2}$. This power of vanishing can be identified with
the power $5/2$ in (\ref{def rho}), and a local parametrix can be
constructed in terms of the RH solution $\Phi$. At least this is
true for values of $x,t,\e$ that are mapped to a complex number $Z$
such that the RH problem for $\Phi$ with parameter $Z$ is solvable
(or in other words values of $Z$ where the tritronqu\'ee solution
$Q$ has no pole). The results in \cite{BT} confirm the conjectured
behavior from \cite{DGK}, but two questions remain unanswered.
First the absence of poles for the tritronqu\'ee solution
$Q(Z)$ to Painlev\'e I in the sector $|\arg Z|<\frac{4\pi}{5}$ has
not been proved yet. Secondly, given initial data for the focusing NLS
equation, semiclassical asymptotics for the reflection coefficient
(which corresponds to a non-self-adjoint Zakharov-Shabat operator)
have not been obtained. In \cite{BT}, an ansatz has been made about
the small $\e$-behavior of the reflection coefficient, and an
asymptotic analysis of the RH problem has been done for such
reflection coefficients. This implies a proof of the conjectured
behavior for initial data that correspond to a reflection
coefficient of the proposed form, but it is not known that the
reflection coefficient corresponding to appropriate $\e$-independent
initial data is of the prescribed form.

\section{Conclusion and outlook}

For Hamiltonian perturbations of hyperbolic equations or systems
which can be 'solved' by inverse scattering, or in other words which
can be characterized in terms of a RH problem, one can hope at this
point to prove that they obey Dubrovins universality conjecture. The
Deift/Zhou steepest descent method allows one to analyze the RH
problem asymptotically in the small dispersion limit $\e\to 0$,
although several of the RH transformations are considerably
different for different equations. In particular the $g$-function
mechanism has to be handled separately for each equation. This
means that it is not obvious at all how one can prove the universality
simultaneously for a whole class of equations. Another issue is that
one needs to have detailed asymptotic information about the
reflection coefficient in the semiclassical limit $\e\to 0$ before
one can start with the asymptotic analysis of the RH problem.
General equations of the form
(\ref{Hampert}) cannot be solved using inverse scattering; for such
'non-integrable' equations there are no rigorous results about the asymptotic behavior available,
even existence of solutions has not been established in general.

In the elliptic case, much less is clearly understood as we
explained in the previous section. Although the tritronqu\'ee
solution to Painlev\'e I has a long history, the conjecture about
its poles remains open. Another challenge is to obtain semiclassical
asymptotics for the reflection coefficient related to the
Zakharov-Shabat equation. Apart from the focusing NLS equation, no
rigorous results have been obtained concerning critical behavior in
the elliptic case.

\section*{Acknowledgements}
The author acknowledges support by the Belgian Interuniversity
Attraction Pole P06/02 and by the ERC program FroM-PDE.

\obeylines \texttt{Tom Claeys
    Universit\'e Catholique de Louvain
Chemin du cyclotron 2
1348 Louvain-La-Neuve
BELGIUM
E-mail:
tom.claeys@uclouvain.be}

\end{document}